\documentclass[11pt,prd,a4paper,preprintnumbers,amsmath,amssymb,nofootinbib]{article}
\pdfoutput=1
\usepackage{feynmp-auto,expdlist}
\usepackage{amsmath,amsfonts,amssymb}
\usepackage{graphicx}
\usepackage{enumerate}
\usepackage{hyperref}
\usepackage{latexsym}
\usepackage{hepnicenames}
\usepackage{enumerate}
\usepackage{soul}
\usepackage[normalem]{ulem}
\usepackage{wasysym}
\usepackage{makecell}
\usepackage{bbm}
\usepackage{physics}
\usepackage[version=4]{mhchem}
\usepackage{amsmath}

\font\tenrsfs=rsfs10 at 12pt
\font\sevenrsfs=rsfs7
\font\fiversfs=rsfs5
\newfam\rsfsfam
\textfont\rsfsfam=\tenrsfs
\scriptfont\rsfsfam=\sevenrsfs
\scriptscriptfont\rsfsfam=\fiversfs

\numberwithin{equation}{section}

\usepackage{mathrsfs}
\usepackage{braket}
\usepackage{titling}
\usepackage{amsmath}
\usepackage{slashed}
\usepackage{amssymb}
\usepackage{epsfig}
\usepackage{graphicx}
\usepackage{color}
\usepackage{rotating}
\usepackage{hyperref}
\usepackage[margin=1.in]{geometry}
\usepackage[table,xcdraw,dvipsnames]{xcolor}
\usepackage[compress,numbers,sort]{natbib}
\usepackage{colortbl}
\usepackage{pdflscape}
\usepackage{color}
\usepackage{mathtools}
\usepackage{colortbl}
\usepackage{comment}
\usepackage{multirow}
\usepackage{booktabs}

\newcommand{\BR}{{\mathcal B}}

\definecolor{nicered}{rgb}{0.7,0.1,0.1}
\definecolor{nicegreen}{rgb}{0.1,0.5,0.1}
\definecolor{red}{rgb}{1.0, 0, 0}
\definecolor{niceblue}{rgb}{0,0,0.8}
\allowdisplaybreaks

\definecolor{blus}{cmyk}{1,1,0,0.6}
\definecolor{verde}{cmyk}{0.92,0,0.59,0.25}
\definecolor{rossos}{cmyk}{0,1,1,0.55}

\hypersetup{colorlinks,bookmarksopen,bookmarksnumbered,linkcolor=blus,pdfstartview=FitH,urlcolor=rossos,citecolor=verde}

\def\eq#1{{Eq.~(\ref{#1})}}

\renewcommand{\bar}{\overline}

\newcommand{\beq}{\begin{equation}}
\newcommand{\eeq}{\end{equation}}
\newcommand{\bea}{\begin{eqnarray}}
\newcommand{\eea}{\end{eqnarray}}

\renewcommand{\[}{\left[}

\renewcommand{\(}{\left(}
\renewcommand{\)}{\right)}

\def\be{\begin{equation}}
\def\ee{\end{equation}}

\begin{document}

\begin{center}  
{\LARGE
\bf\color{blus}
Probing a Light Scalar Boson with a few-MeV \\
\vspace{0.1cm}
Proton Beam Deep Underground
}
\vspace{0.8cm}

{\bf 
Carlo Broggini$^{a}$, 
Giuseppe Di Carlo$^{b}$, 
Luca Di Luzio$^{a}$, \\ 
Denise Piatti$^{c,a}$, 
Claudio Toni$^{d}$ }\\[5mm]

{\it $^a$Istituto Nazionale di Fisica Nucleare, Sezione di Padova, \\
Via F. Marzolo 8, 35131 Padova (PD), Italy}\\[1mm]
{\it $^b$Istituto Nazionale di Fisica Nucleare, Laboratori Nazionali del Gran Sasso, \\
67100 Assergi (AQ), Italy}\\[1mm]
{\it $^c$Dipartimento di Fisica e Astronomia `G.~Galilei', Universit\`a di Padova,
 \\ Via F. Marzolo 8, 35131 Padova (PD), Italy}\\[1mm]
{\it $^d$LAPTh, Université Savoie Mont-Blanc et CNRS,
74941 Annecy, France}

\vspace{0.3cm}
\begin{quote}

We propose to investigate the production of a light scalar boson $\phi$ in low-energy proton-nucleus interactions using the 3.5~MV accelerator of the Bellotti Ion Beam Facility, located in the underground Gran Sasso National Laboratory. Nuclear reactions induced by a few-MeV proton beam on suitable target materials can act as a controlled source of $\phi$ particles. Owing to the deep-underground location, the facility benefits from substantial cosmic-ray shielding, enabling searches for rare processes with minimal background. The produced $\phi$ particles will be sought with large-volume, low-background detectors already operating or currently under construction at the Gran Sasso National Laboratory. This approach combines a tunable accelerator-based production mechanism with the exceptional sensitivity of underground rare-event searches, offering a novel avenue to probe light scalar bosons beyond the Standard Model.

\end{quote}

\thispagestyle{empty}

\end{center}

\setcounter{tocdepth}{2}
\tableofcontents




\section{Introduction}
\label{sec:intro}

Light bosonic particles beyond the Standard Model (SM) are theoretically well-motivated and arise in a variety of extensions of the SM, including models with additional scalar fields 
and dark-sector frameworks. Owing to their feeble couplings, such particles are difficult to probe at high-energy colliders, but may be produced and detected in low-energy laboratory experiments or through rare processes in astrophysics and cosmology. Among these candidates, light scalar bosons with MeV-scale masses provide particularly intriguing targets: they can play a role in dark matter dynamics, impact stellar evolution, and modify precision observables. Dedicated laboratory searches are therefore crucial to test their parameter space in a controlled and systematic way.  

A promising strategy is to exploit nuclear reactions at low energies as a source of new light states, 
employing either underground accelerators or radioactive sources, 
as suggested \emph{e.g.}~in Refs.~\cite{Izaguirre:2014cza,Pospelov:2017kep}. 
Proton-nucleus interactions in the few-MeV regime can efficiently produce scalar particles via nuclear transitions, while the resulting flux is relatively tunable through the choice of beam energy and target material. When combined with large-volume, ultra-low-background detectors, such as those employed in underground rare-event searches, this approach provides a unique opportunity to investigate light scalars with unprecedented sensitivity.  

As a concrete realization of this idea, we consider the 3.5~MV accelerator of the Bellotti Ion Beam Facility (Bellotti-IBF) \cite{SEN2019390,LUNA}, located in the Gran Sasso National Laboratory (LNGS). The underground location ensures substantial shielding from cosmic rays, strongly reducing backgrounds and enabling the study of rare processes. In this setup, proton beams impinging on selected target nuclei act as a controlled source of scalar bosons $\phi$, which can then be searched for with the existing or forthcoming generation of LNGS detectors, such as XENONnT 
\cite{XENON:2024wpa} and DarkSide-20k \cite{DarkSide-20k:2024yfq}.  

The structure of this paper is as follows. In Sec.~\ref{sec:production} we describe the production mechanisms for a light scalar boson in nuclear reactions and provide an estimate of the expected flux. In Sec.~\ref{sec:detection} we discuss detection strategies, focusing on the signatures in large-volume underground detectors. We present the results and outline future prospects in Sec.~\ref{sec:results}, before concluding in Sec.~\ref{sec:concl}. App.~\ref{sec:constraints} is devoted to current constraints on MeV-scale scalar bosons from astrophysics, flavor physics, and laboratory probes.

\section{Production of a light scalar boson}
\label{sec:production}

The 3.5~MV accelerator at the Bellotti-IBF \cite{SEN2019390,LUNA} can deliver proton, helium, and carbon beams \cite{SEN2019390,LUNA}. In this work we focus on the proton beam, which can reach currents up to 1~mA. Proton fusion with a target nucleus ($T$) produces a new nucleus either in its ground state ($N$) or in an excited state ($N_*$). The excited state subsequently de-excites via the standard channels, \emph{i.e.}~gamma or particle emission, or, if the quantum numbers allow, through the emission of a new scalar boson, 
$\phi$.

\subsection{Nuclear production mechanisms}

We can schematically define two classes of photon/scalar production mechanisms:  
\begin{itemize}
\item[1)] Direct nuclear production $p+T\to N+X$;
\item[2)] Nuclear reaction $p+T\to N_* + \dots$ with $N_*\to N+X$,
\end{itemize}
where  $X=\gamma/\phi$.
For simplicity, we will consider only the last production mechanism, which is the easiest to evaluate, thus producing conservative bounds.
We then need
the decay widths for 
\beq
\label{eq:nuc_decay}
\Gamma(N_i \to N_f X) \, ,
\eeq
where $N_{i,f}$ are two energy levels of the same nucleus.
Nuclear states possess definite spin and parity quantum numbers, respectively $J_{i,f}$ and ${\pi_{i,f}}$.
Labeling the total (orbital) angular momentum of the emitted boson as $J$ ($L$) and its parity as $\pi$, conservation laws impose:
\beq
|J_i - J_f |\leq J \leq J_i + J_f \qquad \text{and} \qquad \pi_i = \pi_f \pi (-1)^L \, .
\eeq
If $X$ is a photon, $J\geq1$ and we identify two types of transition:
\begin{itemize}
\item electric type transition $EJ$ if $\pi_i \pi_f=(-1)^{J}$;
\item magnetic type transition $MJ$ if $\pi_i \pi_f=(-1)^{J+1}$.
\end{itemize}
If $X$ is a scalar, one has $J=L\geq 0$ and $\pi_i \pi_f = (-1)^{L}$, corresponding to an electric-type transition, with the exception that a photon cannot be emitted for $J=0$.\footnote{If $X$ is an axion one finds $J=L\geq 0$ and $\pi_i \pi_f = (-1)^{L+1}$, as in a magnetic-type transition, again with the exception that a photon cannot be emitted for $J=0$.}

\subsubsection{Multipole expansion}

We now expand Eq.~\eqref{eq:nuc_decay} in multipoles, with $k$ denoting the boson momentum:
\begin{align}
\Gamma_{\phi} &=\frac{2k}{2J_{i}+1}\Biggl\{ \sum_{J\geq 0}\left|\braket{J_f||\mathcal{G}_{J}||J_i}\right|^{2}\Biggr\} \, , \\
\Gamma_\gamma &=\frac{2k}{2J_{i}+1}\Biggl\{ \sum_{J\geq 1}\left[\left|\braket{J_f|| \mathcal{T}_{J}^\text{el}||J_i}\right|^{2} + \left|\braket{J_f||\mathcal{T}_{J}^\text{mag}||J_i}\right|^{2}\right]\Biggr\} \ ,
\end{align}
where the spherical operators have been introduced 
(see \emph{e.g.}~Ref.~\cite{Barducci:2022lqd} for details and references)
\begin{align}
\mathcal{G}_{JM} & =
\int d^{3}\vec{r}\,j_{J}(kr)Y_{JM}(\hat{r})\mathcal{S}(\vec{r}) \, , \\
\mathcal{T}_{JM}^{\rm el} & =\frac{1}{k}\int d^{3}\vec{r}\,\vec{\nabla}\times[j_{J}(kr)\textbf{Y}_{JJM}(\hat{r})]\cdot\vec{\mathcal{J}}_\gamma(\vec{r}) \, , \\
\mathcal{T}_{JM}^{\rm mag} & =\int d^{3}\vec{r}\,[j_{J}(kr)\textbf{Y}_{JJM}(\hat{r})]\cdot\vec{\mathcal{J}}_\gamma(\vec{r}) \, .
\end{align}
We then consider the nucleon-photon/scalar effective interaction Lagrangians, defined via 
\begin{align}
\mathcal{L}_{\gamma NN}&=A_\mu(x) \mathcal{J}^\mu_\gamma(x)=A^\mu \left[ eQ_{p}\bar{p}\gamma_{\mu}p+eQ_{n}\bar{n}\gamma_{\mu}n+\frac{e\kappa_{p}^{\gamma}}{2m_{p}}\partial^{\nu}(\bar{p}\sigma_{\mu\nu}p)+\frac{e\kappa_{n}^{\gamma}}{2m_{n}}\partial^{\nu}(\bar{n}\sigma_{\mu\nu}n) \right] \, , \\
\mathcal{L}_{\phi NN}&=\phi(x)\mathcal{S}(x)=\phi\, \left[ g_{p}\bar{p}p+g_{n}\bar{n}n\right] \, ,
\end{align} 
where $\kappa_p^\gamma =\mu_p - Q_p= +1.792847351(28)$, $\kappa_n^\gamma =\mu_n - Q_n= -1.9130427(5)$ and $Q_{p,n}$ indicates the electric charge of the nucleon in units of the absolute electron charge. The same methodology can be applied to internal $e^+e^-$ pair creation, see App.~B of Ref.~\cite{Barducci:2022lqd} and references thereby.

Consider the case of proton impinging on a $^{19}$F target. The nuclear fusion will produce $^{20}$Ne which will mainly decay via $\alpha$ emission to $^{16}$O. Among the possible $^{16}$O states populated by $^{20}$Ne $\alpha$-decay, we focus on the $\ce{^{16}O}(6.05)\to\ce{^{16}O}(g.s.)$, where both the initial and final states are $J^\pi=0^+$ nuclei with isospin $I=0$. 
In this case, due to angular momentum conservation, the single-$\gamma$ emission is forbidden and the leading SM decay channel proceeds via internal $e^+e^-$ pair creation, $\ce{^{16}O}(6.05)\to\ce{^{16}O}(g.s.) +e^+e^-$. 
Thus, the relative branching to new physics can
be greatly enhanced \cite{Izaguirre:2014cza}.\footnote{In contrast, in the axion case, the SM background is dominated by the large single-$\gamma$ decay, 
resulting in a strong suppression.} 
Employing isospin symmetry, the scalar emission rate can be related to the internal $e^+e^-$ pair creation rate, and is given by \cite{Resnick:1973vg}\footnote{The calculations of Refs.~\cite{Barducci:2022lqd,Barducci:2025hpg} reproduce the result of Ref.~\cite{Resnick:1973vg} in the full non-relativistic limit of the nuclear current, and further include the next to leading order (NLO) term in such expansion. A naive estimate suggests that the NLO contribution could enhance the predicted branching ratio for scalar emission by a factor of $\sim 10^3$. However, this term depends on the nuclear matrix element of the kinetic operator, which has not been computed in the literature. For this reason, we conservatively neglect this enhancement factor. We also note that Eq.~\eqref{eq:scalaremissionrate} numerically agrees with the result in Ref.~\cite{Pospelov:2017kep}.}
\be
\label{eq:scalaremissionrate}
\BR(\ce{^{16}O}(6.05)\to\ce{^{16}O}+\phi)\approx\frac{15}{8} \left[1-\left(\frac{m_\phi}{6.05\text{ MeV}}\right)^2\right]^{\frac{5}{2}} \, \left(\frac{g_p+g_n}{\alpha}\right)^2 \ .
\ee

\subsection{Estimate of the scalar flux}

The scalar flux from fusion reactions at the Bellotti-IBF receives contributions from all the production mechanisms discussed above. For simplicity, we neglect the direct nuclear contribution, which is difficult to evaluate, and focus instead on scalars produced through resonant reactions followed by subsequent decays. The resulting flux therefore represents a conservative estimate of the total yield. Accordingly, the number of scalars produced in fusion reactions can be approximated as 
\beq
{N}_{\phi} = {N}_\text{POT} \times \sum_{N_*} \left\{ f_{N_*}(T) \times \sum_{N_f} \BR(N_* \to N_f +\phi) \right\} \, ,
\eeq
where $f_{N_*}(T)$ is the multiplicity of the $N_*$ states produced for each proton on target $T$, while ${N}_\text{POT}$ is the number of protons on target (POT) delivered at the Bellotti-IBF.
From this viewpoint, the scalar flux is a superposition of monochromatic components  
\beq
N_{\phi}=\sum_{\omega_{\phi}} N_{\phi}(\omega_{\phi}) \, ,
\eeq
summing over all possible transition energies.
The multiplicity of the excited state can be evaluated as \cite{Izaguirre:2014cza}
\beq
f_{N_*}(T)=n_T \times \int^{E_p}_{0} dE \ \frac{\sigma(p+T\to N_* + \dots )}{|dE/dx|} \, ,
\eeq
where $n_T$ denotes the target density and the stopping power $|dE/dx|$ depends on the target material. 
The ellipsis in the cross section indicates that we perform an inclusive sum over all possible by-products 
($\gamma$, $\alpha$, etc.) accompanying the excited state of interest.

We recall that the final nucleus $N_*$ is produced approximately at rest, so the angular distribution of the scalar flux is entirely given by the nuclear transition and it can be shown that the angular distribution of the emitted boson is isotropic.

On a practical level, we consider a 3~MeV proton beam with an intensity of 1~mA,
equivalent to $6.2\times10^{15}$ POT per second,
impinging on a TaF$_3$ target of about 3~mm thickness. The proton beam will stop inside the target, allowing the cross section to be integrated over the full beam energy.  In this scenario,
the expected multiplicity of the $\ce{^{16}O}(6.05)$ state is found to be:
\beq
f_{\ce{^{16}O}(6.05)}(\text{TaF}_3)= 2.3\times10^{-6} \, .
\eeq
The $N_{*}$ production cross section was taken from the comprehensive R-matrix 
analysis of Ref.~\cite{Lombardo19-PRC}. To be conservative we assumed a vanishing contribution from the 2$^{+}$ ($13095$~keV) resonant state in $^{20}$Ne. The effective stopping power has been calculated using the program SRIM \cite{srim2003}. The profile of the target was assumed to be box-like, with about 10$^{18}$ atoms/cm$^{2}$. It must be noted that compared to calculation reported in \cite{Izaguirre:2014cza}, we used an updated cross section, based on experimental data rather than a simple model, and a realistic thick solid target.


\section{Detection of a light scalar boson}
\label{sec:detection}

The scalar flux emerging from the target can then be intercepted by a suitable detector. 
We consider two scenarios: detection with the existing XENONnT experiment \cite{XENON:2024wpa}, and with the DarkSide-20k setup \cite{DarkSide-20k:2024yfq} currently under construction.

\subsection{Experimental signal signatures}

The experimental signatures of the signal include:
\begin{itemize}
\item a photon from scattering, $\phi + e^- \to e^- + \gamma$;  
\item two photons from the decay, $\phi \to \gamma\gamma$;   
\item an electron-positron pair from the decay, $\phi \to e^+ e^-$.  
\end{itemize}
We consider here also a lepton-scalar coupling through the Lagrangian 
in \eq{eq:Lphitofermions}. 
In our scenario the scalar can only decay into an 
$e^+e^-$ pair or into two photons, thus
\be
\Gamma_\phi=\Gamma(\phi\to\gamma\gamma)+\Gamma(\phi\to e^+e^-) \, .
\ee
In the following, we report the number of expected events  
for each channel.

\subsubsection*{$\phi+e^-\to e^-+\gamma$}

The number of photons produced in the signal volume, with the interaction point set at $\vec{x}_\text{IP}=0$, is given by
\beq
N_\gamma =\sum_{\omega_\phi} N_\phi(\omega_\phi) \int \frac{d\Omega_\phi}{4\pi} \int^\infty_0 dt \ \text{exp}\left\{{-\frac{t}{\tau_\phi \gamma_\phi}+\int_{0}^{t}dt^\prime n_e(\vec{\beta}_\phi t^\prime) \sigma_{\phi\to\gamma}\beta_\phi}\right\} \ n_e(\vec{\beta}_\phi t) \sigma_{\phi\to\gamma}^\text{cut} \beta_\phi \ ,
\eeq
where $\vec{\beta}_\phi$ is the scalar velocity, $\Omega_\phi$ the scalar emission solid angle, 
$\beta_\phi = |\vec{\beta}_\phi| = \sqrt{\omega_\phi^2 - m_\phi^2}/\omega_\phi$, 
$\gamma_\phi = 1/\sqrt{1-\beta_\phi^2}$, 
$\tau_\phi = \Gamma_\phi^{-1}$ the scalar lifetime at rest, 
and $n_e(\vec{x})$ the electron number density of the detector. The cross section reads
\be
\sigma_{\phi\to\gamma}^\text{(cut)} =\left(\frac{g_{e}^2\alpha}{2m_e\omega_\phi^2 \beta_\phi^2}\right) \times \int_\text{(cut)} d\omega_\gamma \  F(m_e^2+m_\phi^2+2m_e\omega_\phi,m_e^2-2m_e\omega_\gamma) \ ,
\ee
with
\be
\begin{split}
F(s,t)=\frac{1}{2}\sum_\text{pol}|{\cal M}(\phi+e^-\to e^-&+\gamma)|^2\\
=-\frac{2}{(m_e^2 - s)^2 (m_e^2 - t)^2} \Big\{ &20 m_e^8 - 4 m_e^6 \left( 3 m_\phi^2 + 4 (s + t) \right) \\
+ &m_e^4 \left( 2 m_\phi^4 + 5 s^2 - 2 s t + 5 t^2 + 10 m_\phi^2 (s + t) \right) \\
- &m_e^2 \left( 4 m_\phi^2 s t + 2 m_\phi^4 (s + t) + (s - t)^2 (s + t) \right) \\
+ &s t \left( 2 m_\phi^4 - 2 m_\phi^2 (s + t) + (s + t)^2 \right)\Big\} \ .
\end{split}
\ee
The minimal and maximal value of the photon energy are
\be
(\omega_\gamma)_\text{min}=\frac{(m_e+\omega_\phi(1-\beta_\phi))(m_\phi^2+2m_e\omega_\phi)}{2(m_e^2+m_\phi^2+2m_e\omega_\phi)} \ , \hspace{5mm} (\omega_\gamma)_\text{max}=\frac{(m_e+\omega_\phi(1+\beta_\phi))(m_\phi^2+2m_e\omega_\phi)}{2(m_e^2+m_\phi^2+2m_e\omega_\phi)}  \ .
\ee
However, to identify the signal, the detector may require veto conditions, thereby modifying the integral into $\sigma_{\phi\to\gamma}^\text{cut}$ to account for this effect. 

In the limit where the scalars arrive at the detector approximately collinearly,  
\emph{i.e.}~when the detector size is much smaller than the distance $L$ between the production 
target and the detector, so that the solid angle of the captured scalar flux is negligible, 
the integration can be simplified and the expression reduces to
\be
\label{eq:sign1}
N_\gamma=\frac{N_\phi}{4\pi L^2} \times V_\text{eff} \times n_e \times \sigma_{\phi\to\gamma}^\text{cut} \times \text{exp}\left\{{-\frac{L}{\tau_\phi \gamma_\phi\beta_\phi}}\right\} \ ,
\ee
which is similar to what has been used in Ref.~\cite{Izaguirre:2014cza}.

\subsubsection*{$\phi\to\gamma\gamma$}

Scalar couplings to charged fermions induce the di-photon decay of the scalar at 1-loop, with a decay rate given by
\be
\label{eq:phitogg}
\Gamma(\phi\to\gamma\gamma)=\frac{\alpha^2m_\phi^3}{512\pi^3}\left|\sum_{f} q^2_f\frac{g_{f}}{m_f} A_{1/2}\left(\frac{m_\phi^2}{4m_f^2}\right)\right|^2 \ ,
\ee
with
\be
A_{1/2}(x)=\frac{2[x+(x-1)f(x)]}{x^2} \ , \hspace{5mm} f(x)=\begin{cases}
\arcsin^2\sqrt{x} &\text{ for $x\leq1$ ,} \\
-\frac{1}{4}\left(\ln\frac{1+\sqrt{1-x^{-1}}}{1-\sqrt{1-x^{-1}}} - i \pi\right)^2  &\text{ for $x>1$ ,}
\end{cases}
\ee
where the sum in \eq{eq:phitogg} is performed over all the fermion fields 
of the low-energy theory
with electric charge $q_f$.
The scalar couplings to nucleons originate at the fundamental level from scalar interactions with quarks and gluons (cf.~discussion in Sec.~\ref{sec:results}).  
To estimate the contribution of the light-quark couplings, we then rely on the calculation of Ref.~\cite{FeruglioLevati}, which incorporates the one-loop effects of pseudoscalar mesons within chiral perturbation theory.

The number of di-photon decays then produced in the detector, assuming that the interaction target is set at the origin, \emph{i.e.}~$\vec{x}_\text{IP}=0$, is given by
\beq
N_{\gamma\gamma} =\sum_{\omega_\phi} N_\phi(\omega_\phi) \int \frac{d\Omega_\phi}{4\pi} \int^\infty_0 dt \ \text{exp}\left\{{-\frac{t}{\tau_\phi \gamma_\phi}+\int_{0}^{t}dt^\prime n_e(\vec{\beta}_\phi t^\prime) \sigma_{\phi\to\gamma}\beta_\phi}\right\} \ \frac{\BR(\phi\to\gamma\gamma)}{\tau_\phi\gamma_\phi} \  \Theta(\vec{\beta}_\phi t) \ ,
\eeq
where
\be
\Theta(\vec{x})=\begin{cases}
1 & \text{if $\vec{x}$ is inside the detector,} \\
0 & \text{if $\vec{x}$ is not inside the detector.}
\end{cases}
\ee
In the limit of $L$ much larger than the detector size, we get
\be
\label{eq:sign2}
N_{\gamma\gamma}=\frac{N_\phi}{4\pi L^2} \times V_\text{eff} \times \frac{\BR(\phi\to\gamma\gamma)}{\tau_\phi \gamma_\phi\beta_\phi} \times \text{exp}\left\{{-\frac{L}{\tau_\phi \gamma_\phi\beta_\phi}}\right\} \ .
\ee

\subsubsection*{$\phi\to e^+e^-$}

For $m_\phi > 2m_e$, Eq.~\eqref{eq:Lphitofermions} gives  
\be
\Gamma(\phi\to e^+e^-)=g_{e}^2\frac{m_\phi}{8\pi}\left(1-\frac{4m_e^2}{m_\phi^2}\right)^{3/2} \ .
\ee
The number of $e^+e^-$ pair decays produced in the detector, assuming that the target is set at the origin, \emph{i.e.}~$\vec{x}_\text{IP}=0$, is then given by
\beq
N_{ee} =\sum_{\omega_\phi} N_\phi(\omega_\phi) \int \frac{d\Omega_\phi}{4\pi} \int^\infty_0 dt \ \text{exp}\left\{{-\frac{t}{\tau_\phi \gamma_\phi}+\int_{0}^{t}dt^\prime n_e(\vec{\beta}_\phi t^\prime) \sigma_{\phi\to\gamma}\beta_\phi}\right\} \ \frac{\BR(\phi\to e^+e^-)}{\tau_\phi\gamma_\phi} \  \Theta(\vec{\beta}_\phi t) \ .
\eeq
In the limit of $L$ much larger than the detector size, we get
\be
\label{eq:sign3}
N_{ee}=\frac{N_\phi}{4\pi L^2} \times V_\text{eff} \times \frac{\BR(\phi\to e^+e^-)}{\tau_\phi \gamma_\phi\beta_\phi} \times \text{exp}\left\{{-\frac{L}{\tau_\phi \gamma_\phi\beta_\phi}}\right\} \ .
\ee

\subsection{Sensitivity of XENONnT and DarkSide-20k}

We emphasize that in the previous calculations we assumed the scalar interactions to occur and be detected inside the central detector of XENONnT, namely the liquid xenon (LXe) time projection chamber (TPC) with an active target mass of 5.9~tonnes \cite{XENON:2024wpa}. We did not include possible interactions in the two larger water Cherenkov vetoes surrounding the cryostat, since their limited energy resolution and high background levels are unlikely to yield a competitive signal-to-background ratio.  

In contrast, for the DarkSide-20k setup 
(under construction) described in \cite{DarkSide-20k:2024yfq}, we account for scalar interactions both in the central liquid argon (LAr) TPC and in the surrounding LAr vetoes, corresponding to a total LAr mass of about 700~tonnes. Unlike water Cherenkov vetoes, the LAr veto operates via scintillation light detection, potentially offering significantly better energy resolution. 

In our analysis we use the detector parameters summarized in Tab.~\ref{tab:det}, and assume
that both XENONnT and DarkSide-20k 
operate with unit detection efficiency.
For the sensitivity projections we require at least 10 signal events per year, assuming an equal number of background events. This benchmark corresponds to the observation of a $2\sigma$ excess: with $N_S+N_B = 20$ total events and $N_B = 10$ background events per year, the statistical uncertainty is $\sqrt{N_S+N_B}\simeq 5$, so that the signal stands at about $2\sigma$ above the background expectation.

To estimate the background rate, we consider the background levels achieved by 
the Borexino experiment at the LNGS
with a 280-ton active-volume liquid-scintillator detector \cite{BOREXINO:2025dbp} (the detector, 
which required ultra-low backgrounds as 
in DarkSide-20k, 
has now been dismounted). In a 145-ton fiducial volume, over 11 years, and after applying cosmogenic temporal and spatial vetoes,  
fewer than 10 background events were collected in a 40 keV energy bin around 6 MeV. A similar background rate will likely be achieved by DarkSide-20k, at least in the central core of the detector.

We finally point out that one of the most important background in accelerator experiments, \emph{i.e.}~the beam-induced backgrounds that produce $\gamma$ rays and neutrons, is absent in our case. First, the energy beam 
is only 3 MeV, and 
the accelerator-target room has 80 cm-thick concrete walls.  In addition, the accelerator-target complex is located in Gran Sasso Hall-B, whereas the DarkSide-20k detector is in Hall-C, with more than 50 m of rock in between.

\begin{table}[t!]
\begin{center}
\begin{tabular}{c|c|c|c}
 & $L$ & $V_\text{eff}$ & $n_e$ \\
\midrule
\midrule
XENONnT\,\cite{XENON:2024wpa} & $60$ m & $2.0$ $\text{m}^3$ & $7.3\times10^{29}$ $\text{m}^{-3}$ \\
\midrule
DarkSide-20k\,\cite{DarkSide-20k:2024yfq} & $90$ m & $500$ $\text{m}^3$  & $3.8\times10^{29}$ $\text{m}^{-3}$  \\
\end{tabular}
\end{center}
\caption{Parameters used in our analysis: distance from the Bellotti-IBF ($L$), active volume ($V_\text{eff}$), and electron number density ($n_e$) for XENONnT and DarkSide-20k.}
\label{tab:det}
\end{table}


\section{Results and future prospects}
\label{sec:results}

In App.~\ref{sec:constraints} we summarize the existing bounds on an MeV-scale 
scalar coupled to nucleons and electrons. These constraints are essential when 
comparing with the future sensitivities of the experimental setup proposed in this work.  

As a validation of our framework, we have verified that our results reproduce those 
of Ref.~\cite{Pospelov:2017kep} when adopting the same simplified model. 
Our assumptions, however, differ in important respects. 
The simplified model of Ref.~\cite{Pospelov:2017kep} was originally motivated by 
the proton radius anomaly \cite{Tucker-Smith:2010wdq} and the muon $g-2$ discrepancy 
\cite{Chen:2015vqy,Batell:2016ove}, both of which are now resolved. 
In particular, explaining the proton radius anomaly required $g_n \ll g_p$, 
while the muon $g-2$ motivated scalar couplings to heavy leptons, which also 
affected the $\phi \to \gamma\gamma$ rate.  

In this work we adopt a simpler setup, assuming couplings only to electrons and 
light quarks at the fundamental level,  
\be
\label{eq:Lphitofermions}
\mathcal{L} \supset \phi \sum_{f=e,u,d,s} g_f \bar{\psi}_f \psi_f \, .
\ee
This choice is further motivated by the strong constraints from 
$K^+ \to \pi^+ \phi$ decay (cf.~App.~\ref{sec:Ktopiphi}), which requires the 
flavor-alignment condition \cite{Delaunay:2025lhl}  
\beq 
\frac{g_u}{m_u} = \frac{g_d}{m_d} = \frac{g_s}{m_s} \equiv \frac{1}{f_\phi} \, ,
\eeq
that we will assume in the following.  

The above condition has direct implications for nucleon couplings.  
The matching between scalar couplings to quarks and nucleons, which is tied to 
nucleon mass generation, reads (see \emph{e.g.}~\cite{Fan:2010gt})
\be
\label{eq:gNgq}
g_N = m_N \sum_{q=u,d,s} f_{Tq}^{(N)} \frac{g_q}{m_q} \, , 
\ee
with $N=p,n$, and neglecting couplings to gluons. 
The nucleon mass fractions are 
\cite{Ellis:2000ds,Cheng:2012qr}
\begin{align}
& f_{Tu}^{(p)}=0.020\pm0.004 \, ,\quad f_{Td}^{(p)}=0.026\pm0.005 \, ,\quad f_{Ts}^{(p)}=0.118\pm0.062 \, ,\\
& f_{Tu}^{(n)}=0.014\pm0.003 \, ,\quad f_{Td}^{(n)}=0.036\pm0.008 \, ,\quad f_{Ts}^{(n)}=0.118\pm0.062 \, . 
\end{align}
Importantly, in the flavor-aligned scenario, the scalar couplings to protons and neutrons cannot 
be tuned independently. Numerically one finds
\beq 
g_p \approx 1.54 \times \left(\frac{100~\text{MeV}}{f_\phi}\right) , \qquad 
g_n \approx 1.58 \times \left(\frac{100~\text{MeV}}{f_\phi}\right) ,
\eeq
so that $g_p \simeq g_n$, up to $\mathcal{O}(3\%)$ isospin-breaking corrections.
The scalar emission rate is then estimated 
from \eq{eq:scalaremissionrate} 
as
\be
\BR(\ce{^{16}O}(6.05)\to\ce{^{16}O}+\phi)\approx1.4\times10^{-9} \, \left(\frac{g_N}{10^{-7}}\right)^2 \, ,
\ee
where we neglected the mass term in the phase space factor and we normalized the scalar-nucleon coupling to a typical benchmark probed 
by the proposed setup.
In contrast to the analysis of Refs.~\cite{Izaguirre:2014cza,Pospelov:2017kep}, which assumed $g_n=0$, 
our framework therefore predicts an enhancement of a factor 4 in the $\phi$ production rate 
(cf.~\eq{eq:scalaremissionrate}) and a factor 2 in the LSND constraint 
(cf.~App.~\ref{sec:LSND}).  
Furthermore, the solar reaction $p+\ce{^{2}H}\to \ce{^{3}He}+\phi$, which produces scalars with 
energy $E_\phi=5.49$ MeV, is proportional to the isovector combination $g_p - g_n$ 
(cf.~\eq{eq:survprobfusion}).  
As a result, constraints from scalar flux detection in 
SNO (via $g_N$) and Borexino (via $g_e$)  
are strongly suppressed in this limit.

\begin{figure}[t!]
\centering
     \includegraphics[width=0.75\textwidth]{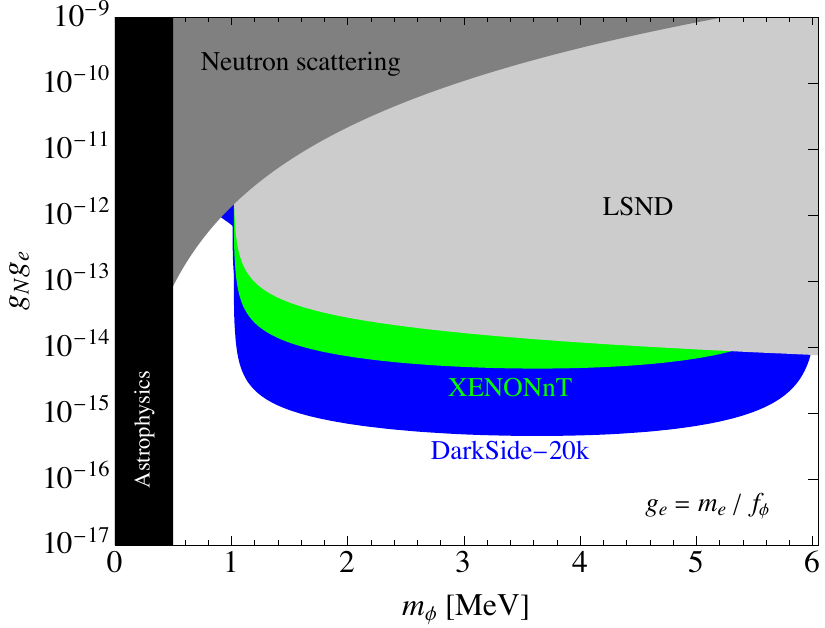}
     \caption{Projected sensitivity of XENONnT (green) and DarkSide-20k (blue), combined with scalar production at the Bellotti-IBF. The projections assume $g_e = m_e / f_\phi$ and correspond to 1 year of data taking with a 3~MeV, 1~mA proton beam on a 3~mm thick TaF$_3$ target. Constraints from LSND (grey), neutron scattering (dark grey), and astrophysics (black) are also shown, while those from the electron $g-2$, SNO and Borexino are subdominant (cf.~App.~\ref{sec:constraints}).}
     \label{fig:prod}
\end{figure}

The projected reach of the setup discussed in this work is shown in Fig.~\ref{fig:prod}, 
where we display the sensitivity of XENONnT (green) and DarkSide-20k (blue) to the product 
coupling $g_N g_e$ as a function of the scalar mass $m_\phi$.  
For comparison, we also show existing bounds from astrophysics (cf.~App.~\ref{sec:astrophysics}), 
neutron scattering (cf.~App.~\ref{sec:neutronscatt}), and the beam-dump experiment LSND (cf.~App.~\ref{sec:LSND}).  
To generate this plot we fixed the electron coupling to $g_e = m_e / f_\phi$.  

\begin{figure}[t!]
\centering
     \includegraphics[width=1.00\textwidth]{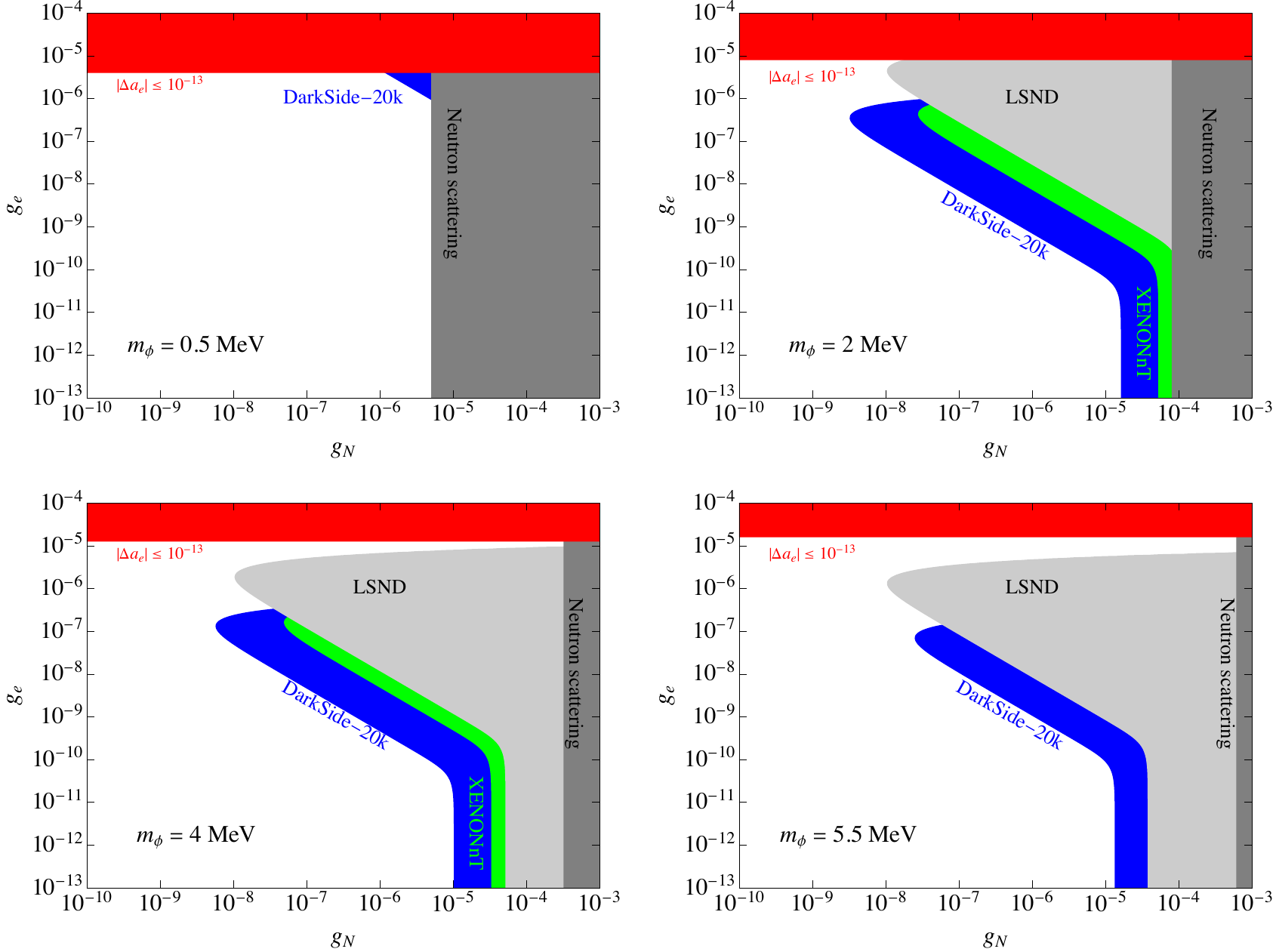}
     \caption{Projected sensitivity of XENONnT (green) and DarkSide-20k (blue) combined with scalar production at the Bellotti-IBF, for fixed values of the scalar mass.  The projections assume a data acquisition time of 1 year with the same beam and target as in Fig.~\ref{fig:prod}. Constraints from LSND (grey), neutron scattering (dark grey), and 
     the electron $g-2$ (red) are also shown, while SNO and Borexino remain subdominant.}
     \label{fig:fixedmass}
\end{figure}

Further insight can be obtained by fixing $m_\phi$ and presenting the reach in the $(g_N, g_e)$ plane.  
This is illustrated in Fig.~\ref{fig:fixedmass}, where we show benchmarks for $m_\phi = (0.5, 2, 4, 5.5)$ MeV.   
In this representation, the anomalous magnetic moment of the electron (cf.~App.~\ref{sec:electrongm2}) provides 
important constraints on $g_e$.  
The sensitivity curves terminate at large $g_e$ values, since in this regime the scalar lifetime becomes too short for $\phi$ to reach the detector.  
A pronounced change also occurs across the $e^+e^-$ threshold at $m_\phi \simeq 1$~MeV, 
where the opening of the efficient $\phi \to e^+ e^-$ detection channel significantly enhances 
the sensitivity. Conversely, as $m_\phi$ approaches the kinematic threshold for scalar production, 
the LSND bounds become dominant over the projected signal reach.  

\begin{figure}[t!]
\centering
     \includegraphics[width=1.00\textwidth]{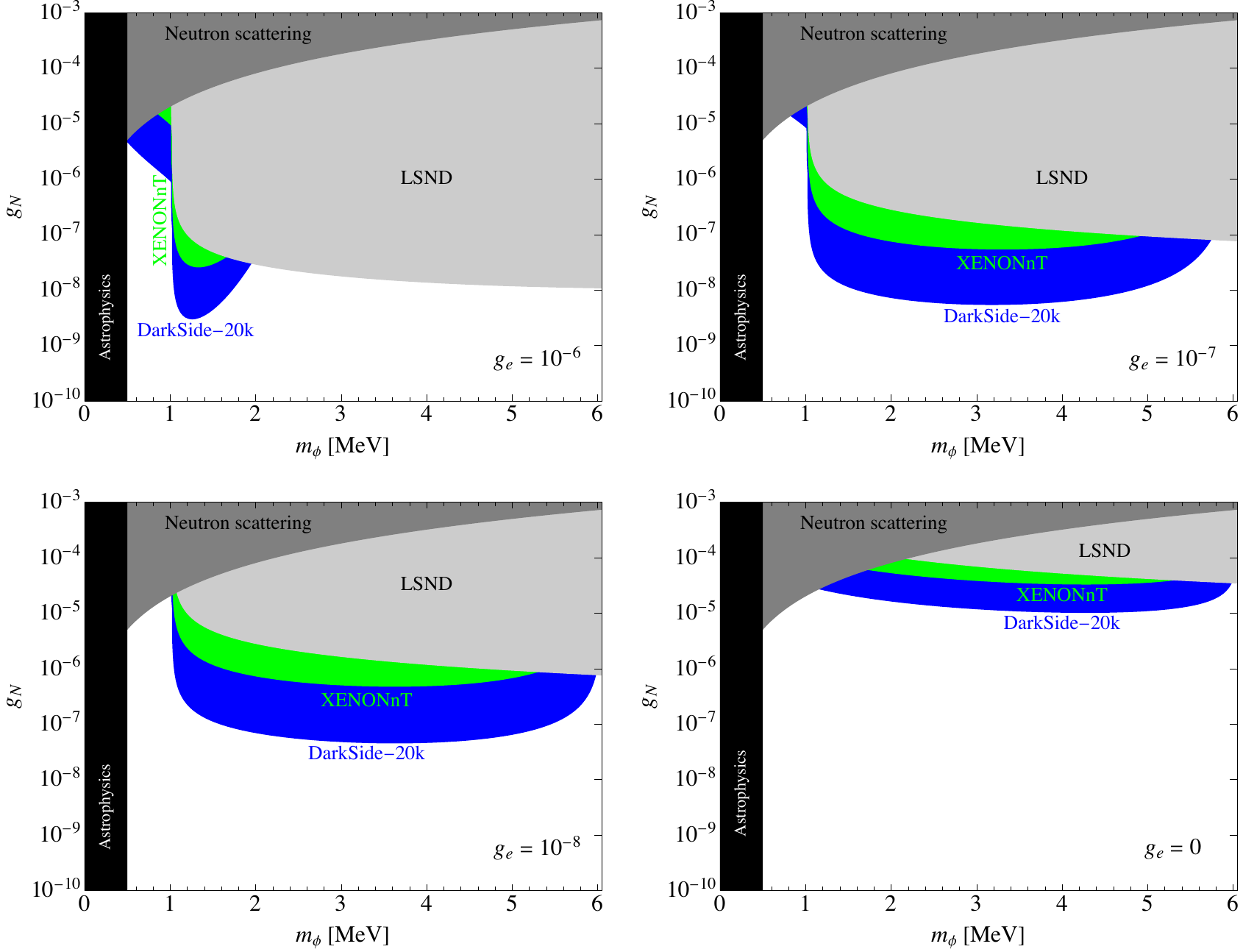}
     \caption{Projected sensitivity of XENONnT (green) and DarkSide-20k (blue) combined with scalar production at the Bellotti-IBF, for fixed values of the electron coupling. The projections assume a data acquisition time of 1 year with the same beam and target as in Fig.~\ref{fig:prod}. Constraints from LSND (grey), neutron scattering (dark grey), and astrophysics (black) are also shown, while SNO and Borexino are subdominant.}
     \label{fig:fixedge}
\end{figure}

Finally, Fig.~\ref{fig:fixedge} presents the complementary case of fixing $g_e$ to benchmark values, 
starting from $g_e \lesssim 10^{-6}$, as allowed by the electron $g-2$.  
This allows one to visualize the reach in terms of the nucleon coupling $g_N$, as a function of $m_\phi$, under different assumptions for the electron coupling.




\section{Conclusions}
\label{sec:concl}

In this work we have proposed and analyzed a novel strategy to search for MeV-scale scalar bosons. 
The central idea is to exploit nuclear reactions induced by low-energy accelerators as controlled 
sources of new light particles, and to search for their flux with large-volume, low-background 
detectors typically designed for rare-event physics. 
As a concrete realization, we have considered the Gran Sasso National Laboratory, where the 3.5~MV 
accelerator at the Bellotti-IBF can serve as a tunable source of nuclear reactions. The resulting scalar 
flux could then be searched for with existing/forthcoming detectors such as XENONnT and DarkSide-20k.

We have discussed the nuclear production mechanisms, focusing on scalar emission from excited 
nuclear states, and provided estimates of the resulting fluxes. We then studied the detection 
channels -- electron scattering, di-photon decays, and $e^+e^-$ decays -- and presented the projected 
sensitivities of XENONnT and DarkSide-20k. These sensitivities were compared with existing bounds 
from astrophysics, neutron scattering, and LSND.  

Our analysis shows that the proposed setup can probe previously unexplored regions of parameter 
space, complementary to astrophysical and laboratory constraints. The approach is parasitic in 
nature, relying on detectors already in operation or construction, and can thus provide a 
cost-effective and innovative probe of light scalar bosons.  

Future work may include a more refined treatment of nuclear production mechanisms, 
as well as
dedicated efficiency, energy resolution and
background studies for the specific detectors. 
Moreover, this proposal could serve as a test bed for probing other light bosonic particles beyond scalars.

\section*{Acknowledgments}

We thank Ferruccio Feruglio, Gabriele Levati and Marco Selvi for useful 
communications. 
The work of LDL is supported
by the European Union -- Next Generation EU and
by the Italian Ministry of University and Research (MUR) 
via the PRIN 2022 project n.~2022K4B58X -- AxionOrigins.
The work of CT has received funding from the French ANR, under contracts ANR-19-CE31-0016 (`GammaRare') and ANR-23-CE31-0018 (`InvISYble'), that he gratefully acknowledges.

\appendix


\section{Constraints on MeV-scale scalar bosons}
\label{sec:constraints}

In this Appendix, we collect present bounds on light scalar bosons 
coupled to electrons and nucleons, specifically for $m_\phi \lesssim 6$ MeV, that is relevant for the new search discussed in this work. 

\subsection{Astrophysical constraints}
\label{sec:astrophysics}

Thermal production of light scalar bosons may lead to energy loss in stars, thus providing strong constraints on the 
scalar coupling to electrons $(g_e)$ and nucleons $(g_N)$, see \emph{e.g.}~\cite{Hardy:2016kme,Bottaro:2023gep,Hardy:2024gwy,Fiorillo:2025zzx}. 
These bounds are exponentially suppressed if the mass of the scalar boson is larger than the typical temperature of the astrophysical object. 
Therefore, we focus on the region $m_\phi \gtrsim 0.5$~MeV, 
where the constraints on $g_e$ from horizontal-branch stars, red giants, 
and white-dwarf cooling become negligible, 
and the leading astrophysical limits on $g_N$ instead originate from neutron-star cooling \cite{Fiorillo:2025zzx} and SN 1987A \cite{Hardy:2024gwy}.  
However, we do not display explicit astrophysical bounds in our plots, 
as existing studies typically consider one coupling at a time, 
while the simultaneous presence of multiple interactions could change the picture. 
In the case of supernovae, constraints can be very strong in the free-streaming regime, 
but the trapping regime is affected by sizable uncertainties. 
In particular, the reabsorption of beyond the SM particles inside the proto-neutron star 
can influence multidimensional processes such as convection and accretion, 
whose impact on the neutrino signal is not yet reliably quantified. 
For these reasons, we simply refer the reader to the discussion in Ref.~\cite{Hardy:2024gwy}, 
emphasizing these caveats.

\subsection{$K^+\to\pi^+\phi$}
\label{sec:Ktopiphi}

Flavor changing Kaon decays involving a light particle 
with a missing energy signature are well constrained by collider searches at NA62, namely ${\cal B}(K^+\to\pi^+\phi)\lesssim{\cal O}(10^{-11})$~\cite{NA62:2024pjp,Guadagnoli:2025xnt,NA62:2025upx}. Even if the scalar couplings are all diagonal in flavor space, such decay can be induced by a $W$-loop. 
For a recent calculation, including next-to-leading order 
corrections, see Ref.~\cite{Delaunay:2025lhl}. 
Using these results, one finds 
that if the light scalar couples only to light quarks 
(\emph{i.e.}~not to gluons) through the Lagrangian term in 
\eq{eq:Lphitofermions},
and if these couplings are proportional to the quark masses, 
$g_{u}/m_u = g_{d}/m_d = g_{s}/m_s\equiv1/f_\phi$, then the leading-order expression for 
$K^+ \to \pi^+ \phi$ in the chiral expansion vanishes. 
In this flavor-aligned scenario, the constraint on $g_{N}$ from $K^+ \to \pi^+ \phi$ searches 
is significantly weakened\footnote{For comparison, in the case of a Higgs-mixed light scalar, the bound from $K^+ \to \pi^+ \phi$ 
corresponds to $g_{N} \lesssim 10^{-7}$~\cite{Baruch:2025lbw}.} and becomes subleading compared to other bounds discussed below.

\subsection{Neutron scattering experiments}
\label{sec:neutronscatt}

As discussed in Ref.~\cite{Tucker-Smith:2010wdq}, neutron scattering experiments provide stringent bounds on new light mediators that couple to neutrons. In particular, a mediator with mass in the MeV range induces corrections to the neutron-nucleus scattering cross section, which can interfere with the strong interaction amplitude. The resulting angular distortions in the differential cross section allow for bounds on the neutron coupling strength, originally derived in Ref.~\cite{Barbieri:1975xy} (see also \cite{Schmiedmayer:1991zz,Leeb:1992qf}). For a mediator with equal couplings to protons and neutrons, 
in the relevant mass range considered in this work, 
this leads to 
\beq 
g_N \lesssim 2 \times 10^{-5} \left( \frac{m_\phi}{\text{MeV}}\right)^2 \, . 
\eeq
In our case, since we do not require the mediator to address the muon $g-2$ or muonic hydrogen anomalies (as \emph{e.g.}~in \cite{Tucker-Smith:2010wdq}), couplings of comparable strength to protons and neutrons remain consistent with current bounds, while still being testable in the parameter space accessible to the setup proposed in this work.

\subsection{LSND}
\label{sec:LSND}

The LSND measurements of the elastic electron-neutrino cross section~\cite{LSND:2001aii,LSND:2001akn} 
can be reinterpreted as constraints on light scalar particles.  
A full recast would require a dedicated analysis; here, as a first approximation, we follow the approach of Ref.~\cite{Pospelov:2017kep}.

Assuming that the decay of the $\Delta$ resonance saturates the pion production inside the target, one can estimates the scalar production from the decay $\Delta \to N+\pi+\phi$, with the scalar emitted from the nucleon leg, as
\be
N_\phi^\text{LSND} \approx N_\pi^\text{LSND} \times \sum_{N=p,n}\frac{\Gamma(\Delta \to N+\pi+\phi)}{\Gamma(\Delta \to N+\pi)} \times P_\text{signal} \, .
\ee
The last term in the above expression accounts for the probability that the emitted scalar produces a signal inside the detector, given by
\be
P_\text{signal}=\Bigg[\text{exp}\left\{-\frac{L_\text{LSND}+\frac{d_\text{LSND}}{2}}{\tau_\phi \gamma_\phi\beta_\phi}\right\}-\text{exp}\left\{-\frac{L_\text{LSND}-\frac{d_\text{LSND}}{2}}{\tau_\phi \gamma_\phi\beta_\phi}\right\}\Bigg]\left(\frac{A_\text{LSND}}{4\pi L^2_\text{LSND}}\right) \, ,
\ee
where $L_\text{LSND}=30$ m, $d_\text{LSND}=8.3$ m and $A_\text{LSND}\approx25$ m$^2$.
Relying again on Ref.~\cite{Pospelov:2017kep}, we consider the approximated result
\be
\frac{\Gamma(\Delta \to N+\pi+\phi)}{\Gamma(\Delta \to N+\pi)} \approx 0.04 \times g^2_N \, ,
\ee
with the average energy of the scalar estimated as 300 MeV.
Finally, a conservative estimate of the number of pions produced in the experiment is $N_\pi^\text{LSND}\sim10^{22}$, while the number of signal events is taken to be less than 20 as in Ref.~\cite{Pospelov:2017kep}.

\subsection{Solar $\phi$ flux from $p+\ce{^{2}H}$ fusion}
\label{sec:solarflux}

Light scalars can be produced via nuclear reactions occurring in the Sun, and later be detected on Earth.
We focus here on the reaction $p+\ce{^{2}H}\to \ce{^{3}He}+\phi$, which generates a flux of scalars with energy of $5.49$ MeV.

In the Sun, deuterium ($\ce{^{2}H}$) is produced 99.6\% of cases through the proton-proton chain, $p+p\to \ce{^{2}H} + e^+ + \nu$, whose neutrino flux $\Phi_{pp\nu}\approx 6\times10^{10} \, \text{cm}^{-2} \, \text{s}^{-1}$ has been measured by Borexino \cite{BOREXINO:2018ohr}, while the remaining 0.4\% is due to $p+p+e^-\to \ce{^{2}H} + \nu$. The deuterium is immediately converted into Helium-3 via $p+ \ce{^{2}H} \to \ce{^{3}He} + \gamma$, or, in our scenario, also via scalar boson emission (in place of a photon), though with a suppressed rate. The scalar flux can be thus estimated
by appropriately rescaling the $\Phi_{pp\nu}$ flux 
(see also \cite{Pospelov:2017kep})
\be
\label{eq:solarflux}
\Phi_{\phi}^\text{Sun}\approx \frac{\sigma(p+ \ce{^{2}H} \to \ce{^{3}He} + \phi)}{\sigma(p+ \ce{^{2}H} \to \ce{^{3}He} + \gamma)}\times\Phi_{pp\nu} \times P_\text{exit} \times P_\text{survive} \, ,
\ee
where we have also included terms accounting for the probability of $\phi$ escaping the star and reaching the Earth. The former is given by
\be
P_\text{exit}=\text{exp}\left\{-\int^{R_\odot}dr \, n_\odot (r)\times\sigma_{\phi\to\gamma}\right\} \, ,
\ee
with $R_\odot=6.96\times10^{10}$ cm 
denoting the solar radius, 
and $n_\odot$ the mean solar electron density, that is approximately described by the exponential function \cite{Bahcall:2000nu}
\be
\frac{n_\odot(r)}{N_A}\approx 245 \, \text{exp}\left\{-10.54 \frac{r}{R_\odot}\right\} \text{  cm}^{-3} \, ,
\ee
where $N_A=6.022\times10^{23}$ is the Avogadro's number.
The survival probability until the Earth is instead given by
\be
\label{eq:survprobfusion}
P_\text{survive}=\text{exp}\left\{{-\frac{L_\odot}{\tau_\phi \gamma_\phi\beta_\phi}}\right\} \, ,
\ee
where $L_\odot=1.5\times10^{11}$ m is the Sun-Earth distance.

The cross section of the $\gamma$-emission is a sum of M1 (s-wave) and E1 (p-wave) multipole contributions with a ratio $\sigma_\text{M1}/\sigma_\text{E1}\approx1.3/2.9$~\cite{Raffelt:1982dr}, with both of them predominately isovectorial~\cite{Eichmann:1963}.
The scalar emission cross section can then be obtained by rescaling the E1 contribution of the electromagnetic one yielding \cite{Pospelov:2017kep,Barducci:2025hpg}
\be
\frac{\sigma(p+ \ce{^{2}H} \to \ce{^{3}He} + \phi)}{\sigma(p+ \ce{^{2}H} \to \ce{^{3}He} + \gamma)} = 
\frac{2.9}{1.3+2.9} \times \frac{1}{2} \, \left[1-\left(\frac{m_\phi}{5.49\text{ MeV}}\right)^2\right]^{\frac{3}{2}} \,
\left(\frac{g_p-g_n}{e}\right)^2 \, .
\ee

\subsubsection{SNO (detection via $g_N$)}
\label{sec:SNObound}

A significant constraint on light scalar interactions with nucleons can be derived by recasting the Sudbury Neutrino Observatory (SNO) experiment \cite{SNO:2011hxd} bound originally set for axion-like particles (ALPs) \cite{Bhusal:2020bvx}, as discussed in Ref.~\cite{Baruch:2025lbw}. This bound relies on the assumption that the new particles produced in solar nuclear transitions are sufficiently long-lived to reach the Earth-based SNO detector, where they could be observed via deuterium dissociation events. 

In analogy to the ALP case, we reinterpret the SNO bound in the case of a light scalar boson $\phi$ that couples to nucleons. The scalar is assumed to be emitted from nuclear transitions in the Sun and then detected via its interaction with deuterium in the SNO detector, provided it survives the propagation from the solar core to Earth. The scalar flux from the Sun is taken to be the one in \eq{eq:solarflux}.
The detection cross section is given by Eq.~(4.3) in 
Ref.~\cite{Baruch:2025lbw}, which describes deuterium dissociation induced by scalar absorption.

This bound becomes ineffective in regions of parameter space where the scalar decays before reaching the Earth. Nevertheless, in the regime where the scalar is sufficiently long-lived, the recast SNO data provide a relevant and competitive constraint on light scalar couplings to nucleons.

\subsubsection{Borexino (detection via $g_e$)}
\label{sec:BOREXINObound}

Similarly to SNO, the scalar flux on Earth is detectable at Borexino if the particle is long-lived, see Ref.~\cite{Borexino:2012guz}. The expected number of $\phi+e^- \to e^- + \gamma$ events is constrained as
\be
\Phi_{\phi}^\text{Sun} \times \sigma_{\phi\to\gamma} \times n_e^\text{Bor.} \times T^\text{Bor.} \times \epsilon_\text{eff}^\text{Bor.} <6.9 \text{ at 90\% C.L.} \, ,
\ee
where $n_e^\text{Bor.}=9.17\times10^{31}$ is the number of electrons
$T^\text{Bor.}=4.63\times10^7$ s is the exposure time and $\epsilon_\text{eff}^\text{Bor.}=0.358$ is the efficiency.
Furthermore the expected number of $\phi\to\gamma\gamma$ decay inside the Borexino detector is constrained as
\be
\Phi_{\phi}^\text{Sun} \times V_\text{eff}^\text{Bor.} \times \frac{\BR(\phi\to \gamma\gamma)}{\tau_\phi \gamma_\phi\beta_\phi} <8.4 \text{ at 90\% C.L.} \, ,
\ee
where $V_\text{eff}^\text{Bor.}$ is the Borexino active volume described as a sphere of radius $R=3.02$ m.

\subsection{Anomalous magnetic moment of the electron}
\label{sec:electrongm2}

The anomalous magnetic moment of the electron, $a_e \equiv (g_e-2)/2$, has been commonly used to extract the value of the fine-structure constant, $\alpha$. However, recent improvements in atomic-physics experiments using Cesium (Cs) and Rubidium (Rb) interferometry have led to the following results for $\alpha$:
\begin{align}
\alpha({\rm Cs}) &= 1/137.035999046(27) \quad \text{\cite{Parker:2018vye}} \, , \\
\alpha({\rm Rb}) &= 1/137.035999206(11) \quad \text{\cite{Morel:2020dww}} \,,
\end{align}
showing a disagreement of $5.5\, \sigma$. Using the above determinations of $\alpha$ to predict the SM value $a^{\rm SM}_e$ and comparing it with the latest experimental measurement of $a^{\rm exp}_e = (115\,965\,218\,059~\pm~13) \times 10^{-14}$~\cite{Fan:2022eto}, yields
the following values of $\Delta a_e \equiv a_e^{\rm exp} - a_e^{\rm SM}$:
\begin{align}
\label{eq:daeCs} 
(\Delta a_e)_{\rm Cs} &= (-102 \pm 26)\times 10^{-14} 
\, , \\
\label{eq:daeRb}
(\Delta a_e)_{\rm Rb} &= (34 \pm 16)\times 10^{-14}
\,.
\end{align}
The contribution of the scalar $\phi$,  
stemming from the electron coupling defined in Eq.~(\ref{eq:Lphitofermions}), reads~\cite{Jegerlehner:2017gek}
\begin{align}
\label{eq:Deltaaephi}
\Delta a^\phi_e &= 
\frac{g^2_{e}}{4\pi^2} \frac{m^2_e}{m^2_\phi} \frac{1}{2} 
\int_0^1 dx \frac{x^2(2-x)}{1-x+\frac{m_e^2}{m_\phi^2} x^2}
\nonumber \\
& 
\approx 
\frac{g^2_{e}}{4\pi^2} \frac{m^2_e}{m^2_\phi} \( \ln\frac{m_\phi}{m_e} -\frac{7}{12} \) \, ,   
\end{align}
where the approximation in the last step is valid for
$m_\phi \gg m_e$. 
Using the latter expression, we obtain the following numerical estimate
\beq
\Delta a^\phi_e \approx 1.3 \times 10^{-13} \left(\frac{g_{e}}{10^{-5}}\right)^{2}
\left(\frac{2\,{\rm MeV}}{m_\phi}\right)^{2}\, ,  
\label{eq:gm2_predictionS}
\eeq
which, for $m_\phi = 2~\text{MeV}$, approximates the exact result at the $20\%$ level. 
In our numerical analysis, however, we employ the full expression given in the first line of Eq.~\eqref{eq:Deltaaephi}.

Here, one could assume three benchmark scenarios: 
$i)$ $|\Delta a_e| \leq 10^{-12}$, where we inflated the current experimental errors on $a^{\alpha({\rm Cs})}_e$ and 
$a^{\alpha({\rm Rb})}_e$ to make Eqs.~(\ref{eq:daeCs}) and 
(\ref{eq:daeRb}) consistent, 
$ii)$ $|\Delta a_e| \leq 10^{-13}$,
assuming a resolution of the current discrepancy in the measurements 
of $\alpha$ with a precision of $\mathcal{O}(10^{-13})$, and 
$iii)$ 
$|\Delta a_e| \leq 10^{-14}$, which is the ultimate expected uncertainty on $\Delta a_e$ if both the errors on 
$a^{\rm exp}_e$ and 
$a^\alpha_e$ will improve by roughly one order of magnitude~\cite{DiLuzio:2024sps}. In practice, we adopt scenario $ii)$ for our analysis, noting that the results for the other cases can be obtained by a straightforward rescaling.  

\begin{small}

\bibliographystyle{utphys}
\bibliography{bibliography.bib}

\end{small}

\end{document}